\documentclass[twocolumn,twocolumnappendix]{aastex631}
\shorttitle{Planetary hydrogen-line scaling relations}  %
\shortauthors{Marleau \&\ Aoyama}

\graphicspath{{./}{Abb/}}  %

\usepackage{etoolbox}

\makeatletter

\patchcmd{\NAT@citex}
  {\@citea\NAT@hyper@{%
     \NAT@nmfmt{\NAT@nm}%
     \hyper@natlinkbreak{\NAT@aysep\NAT@spacechar}{\@citeb\@extra@b@citeb}%
     \NAT@date}}
  {\@citea\NAT@nmfmt{\NAT@nm}%
   \NAT@aysep\NAT@spacechar\NAT@hyper@{\NAT@date}}{}{}

\patchcmd{\NAT@citex}
  {\@citea\NAT@hyper@{%
     \NAT@nmfmt{\NAT@nm}%
     \hyper@natlinkbreak{\NAT@spacechar\NAT@@open\if*#1*\else#1\NAT@spacechar\fi}%
       {\@citeb\@extra@b@citeb}%
     \NAT@date}}
  {\@citea\NAT@nmfmt{\NAT@nm}%
   \NAT@spacechar\NAT@@open\if*#1*\else#1\NAT@spacechar\fi\NAT@hyper@{\NAT@date}}
  {}{}

\makeatother
\makeatletter
\let\jnl@style=\rm
\def\ref@jnl#1{{\jnl@style#1}}

\def\aj{\ref@jnl{AJ}}                   %
\def\actaa{\ref@jnl{Acta Astron.}}      %
\def\araa{\ref@jnl{ARA\&A}}             %
\def\apj{\ref@jnl{ApJ}}                 %
\def\apjl{\ref@jnl{ApJ}}                %
\def\apjs{\ref@jnl{ApJS}}               %
\def\ao{\ref@jnl{Appl.~Opt.}}           %
\def\apss{\ref@jnl{Ap\&SS}}             %
\def\aap{\ref@jnl{A\&A}}                %
\def\aapr{\ref@jnl{A\&A~Rev.}}          %
\def\aaps{\ref@jnl{A\&AS}}              %
\def\azh{\ref@jnl{AZh}}                 %
\def\baas{\ref@jnl{BAAS}}               %
\def\bac{\ref@jnl{Bull. astr. Inst. Czechosl.}}
\def\caa{\ref@jnl{Chinese Astron. Astrophys.}}
\def\cjaa{\ref@jnl{Chinese J. Astron. Astrophys.}}
\def\icarus{\ref@jnl{Icarus}}           %
\def\jcap{\ref@jnl{J. Cosmology Astropart. Phys.}}
\def\jrasc{\ref@jnl{JRASC}}             %
\def\memras{\ref@jnl{MmRAS}}            %
\def\mnras{\ref@jnl{MNRAS}}             %
\def\na{\ref@jnl{New A}}                %
\def\nar{\ref@jnl{New A Rev.}}          %
\def\pra{\ref@jnl{Phys.~Rev.~A}}        %
\def\prb{\ref@jnl{Phys.~Rev.~B}}        %
\def\prc{\ref@jnl{Phys.~Rev.~C}}        %
\def\prd{\ref@jnl{Phys.~Rev.~D}}        %
\def\pre{\ref@jnl{Phys.~Rev.~E}}        %
\def\prl{\ref@jnl{Phys.~Rev.~Lett.}}    %
\def\pasa{\ref@jnl{PASA}}               %
\def\pasp{\ref@jnl{PASP}}               %
\def\pasj{\ref@jnl{PASJ}}               %
\def\rmxaa{\ref@jnl{Rev. Mexicana Astron. Astrofis.}}%
\def\qjras{\ref@jnl{QJRAS}}             %
\def\skytel{\ref@jnl{S\&T}}             %
\def\solphys{\ref@jnl{Sol.~Phys.}}      %
\def\sovast{\ref@jnl{Soviet~Ast.}}      %
\def\ssr{\ref@jnl{Space~Sci.~Rev.}}     %
\def\zap{\ref@jnl{ZAp}}                 %
\def\nat{\ref@jnl{Nature}}              %
\def\iaucirc{\ref@jnl{IAU~Circ.}}       %
\def\aplett{\ref@jnl{Astrophys.~Lett.}} %
\def\apspr{\ref@jnl{Astrophys.~Space~Phys.~Res.}}
\def\bain{\ref@jnl{Bull.~Astron.~Inst.~Netherlands}} 
\def\fcp{\ref@jnl{Fund.~Cosmic~Phys.}}  %
\def\gca{\ref@jnl{Geochim.~Cosmochim.~Acta}}   %
\def\grl{\ref@jnl{Geophys.~Res.~Lett.}} %
\def\jcp{\ref@jnl{J.~Chem.~Phys.}}      %
\def\jgr{\ref@jnl{J.~Geophys.~Res.}}    %
\def\jqsrt{\ref@jnl{J.~Quant.~Spec.~Radiat.~Transf.}}
\def\memsai{\ref@jnl{Mem.~Soc.~Astron.~Italiana}}
\def\nphysa{\ref@jnl{Nucl.~Phys.~A}}   %
\def\physrep{\ref@jnl{Phys.~Rep.}}   %
\def\physscr{\ref@jnl{Phys.~Scr}}   %
\def\planss{\ref@jnl{Planet.~Space~Sci.}}   %
\def\procspie{\ref@jnl{Proc.~SPIE}}   %

\def\ptp{\ref@jnl{Prog.~Th.~Phys.}}   %

\makeatother

\defcitealias{goli04}{G04}
\defcitealias{marl07}{M07}
\defcitealias{scvh}{SCvH}
\defcitealias{bl94}{BL94}
\defcitealias{mc14}{MC14}
\defcitealias{ensman94}{E94}
\defcitealias{lever81}{LP81}

\defcitealias{m16Schock}{Paper~I}
\defcitealias{m18Schock}{Paper~II}

\defcitealias{sm18}{SM18}
\defcitealias{sm19}{SM19}

\defcitealias{AMIM21L}{Ao21}

\usepackage{xspace}

\usepackage{amsmath,amssymb,listings,upgreek,nicefrac}
\usepackage{mathptmx}   %
\usepackage{grffile}
\usepackage{xcolor,xspace}
\usepackage[normalem]{ulem}  %

\usepackage[T1]{fontenc}
\usepackage{textcomp}     %

\newcommand{\vbsrt}[1]{#1}                         %

\newcommand{\auskommentiert}[1]{}                  %
\def\e{\mathrm{e}}                                 %
\def\LSonne{\ensuremath{{L_\odot}}\xspace}         %

\def\Ha{\ensuremath{\mathrm{H}\,\alpha}\xspace}           %
\def\PDSb{PDS\,70\,b\xspace}                              %
\def\Dlrmb{Delorme\,1\,(AB)b\xspace}                      %

\def\MPkt{\ensuremath{\dot{M}}\xspace}                           %

\newcommand{\Lacc}{\ensuremath{{L_{\textnormal{acc}}}}\xspace}  %

\def\LHa{\ensuremath{L_{\textnormal{H}\,\alpha}}\xspace}    %
\def\Lline{\ensuremath{L_{\mathrm{line}}}\xspace}  %

\def\Mdot{\MPkt}

\def\Lsun{\LSonne}

\def\ni{\ensuremath{n_{\mathrm{i}}}\xspace}
\def\nf{\ensuremath{n_{\mathrm{f}}}\xspace}

\renewcommand{\arraystretch}{1.4}

\begin{document}

\title{Planetary line-to-accretion luminosity scaling relations: Extrapolating to higher-order hydrogen lines}

\author[0000-0002-2919-7500]{Gabriel-Dominique Marleau}
\affiliation{
Fakult\"at f\"ur Physik,
Universit\"at Duisburg-Essen,
Lotharstra\ss{}e 1,
47057 Duisburg, Germany%
}
\affiliation{%
Institut f\"ur Astronomie und Astrophysik,
Universit\"at T\"ubingen,
Auf der Morgenstelle 10,
72076 T\"ubingen, Germany%
}
\affiliation{%
Physikalisches Institut,
Universit\"{a}t Bern,
Gesellschaftsstr.~6,
3012 Bern, Switzerland%
}
\affiliation{%
Max-Planck-Institut f\"ur Astronomie,
K\"onigstuhl 17,
69117 Heidelberg, Germany%
}
\email{gabriel.marleau@uni-tuebingen.de, yaoyama@pku.edu.cn}

\author[0000-0003-0568-9225]{Yuhiko Aoyama}
\affiliation{%
Kavli Institude for Astronomy and Astrophysics,
Peking University,
Beijing 100871, People's Republic of China%
}
\affiliation{%
Institute for Advanced Study,
Tsinghua University,
Beijing 100084, People's Republic of China%
}

%

%
%
%
\published{9 December 2022 at RNAAS}

\begin{abstract}
\citet{AMIM21L}
provided scaling relations between hydrogen-line luminosities and the accretion luminosity for planetary-mass objects.
These fits should be an improvement over blind extrapolations of stellar relations.
The fits %
go up only to the $n=8$ electron energy level, but higher-$n$ Balmer lines have been observed in the near-UV \vbsrt{at Delorme~1 (AB)b} with UVES \vbsrt{\citep{ringqvist23}}.
We extend the scaling relations to higher-$n$ levels for the Balmer and other series %
by fitting the fit coefficients $(a,b)$ themselves
and extrapolating them.
Within the assumption of an accretion shock as the source of line emission, these fits should be robust for accreting planetary-mass objects.
\end{abstract}

\section{Motivation}

Planetary-mass companions have been detected or observed at emission lines (e.g., %
\citealp{Haffert+2019,wu22,betti22c,betti22b,ringqvist23}). As for Classical T Tauri Stars (CTTSs),
the lines trace the accretion process, and should originate from an accretion shock, magnetospheric accretion columns, or both (\citealp{AMIM21L}; hereafter \citetalias{AMIM21L}).
Empirically, each line luminosity \Lline correlates with total accretion luminosity \Lacc, which is set mostly by the accretion rate \Mdot. Thus, emission-line observations yield constraints on the formation mechanism and formation timescale of the accretors.

Different models attempt to convert emission-line luminosities
into an accretion rate
(\citealp{aoyama18,Aoyama+2020,Aoyama+Ikoma2019}; \citetalias{AMIM21L}; \citealp{thanathibodee19,szul20,dong21}). %
A popular approach extrapolates the empirical \Lacc--\Lline correlation for CTTSs \citep[e.g.,][]{rigliaco12,alcal17,Komarova+Fischer2020} down to planetary masses.  %
In \citetalias{AMIM21L}, we pointed out
that at \Ha luminosities $\LHa\lesssim10^{-6}~\LSonne$ (near the \LHa of \PDSb; \citealp{zhou21}), extrapolating \citet{rigliaco12} predicts more \Ha emission than incoming energy. %
Thus, blind extrapolations might be invalid.

In \citetalias{AMIM21L} we gave \Lline--\Lacc scaling relations meant to apply at planetary masses. They are based on a simplified model of the accretion geometry but combined with detailed, NLTE hydrogen-line emission calculations \citep{aoyama18}.
We computed and fit \Lacc only up to the $n=8$ energy level since the %
microphysical model follows electron populations only up to a certain level.

UVES observations of \Dlrmb \citep{ringqvist23} motivate us to extend the relationships to higher $n$, thinking also of CUBES \citep{alcal22} or NIR instruments.
We \textit{fit the fit coefficients} %
of our existing relations
and extrapolate them to higher-order lines.
\section{Fit of the fit coefficients, and their extrapolation}

\def\Hoehe{1.73in}
\begin{figure*}[!ht]
\begin{center}
    \includegraphics[height=\Hoehe]{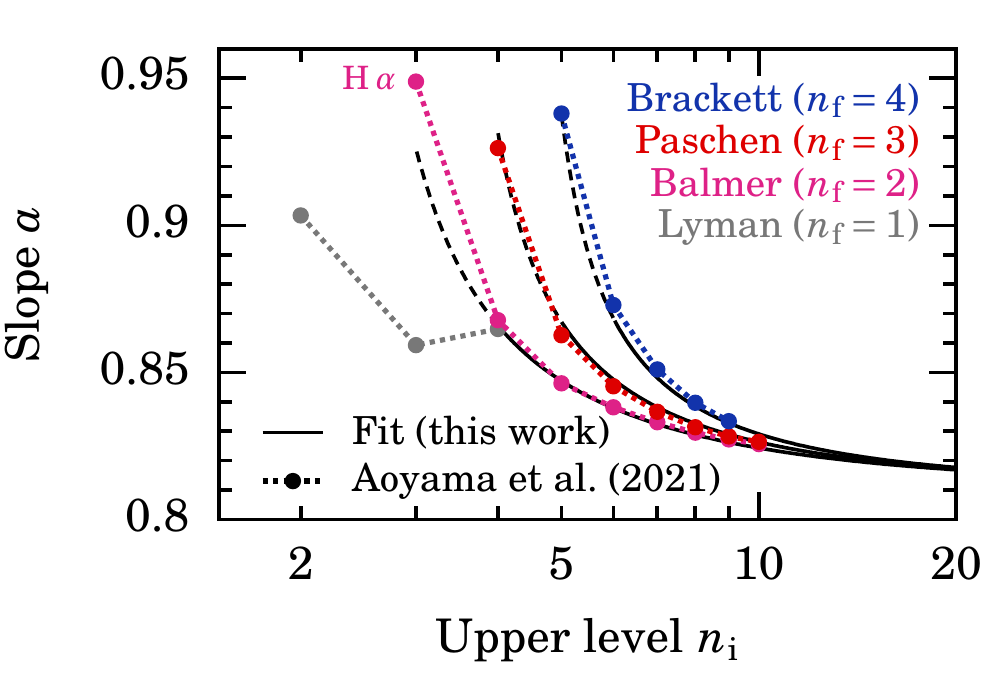}
    \includegraphics[height=\Hoehe]{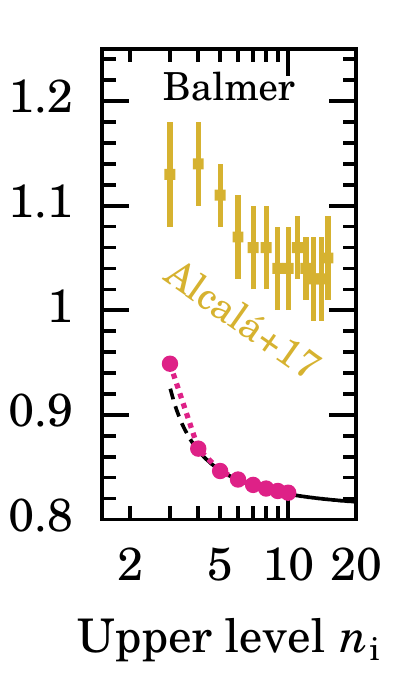} %
    \includegraphics[height=\Hoehe]{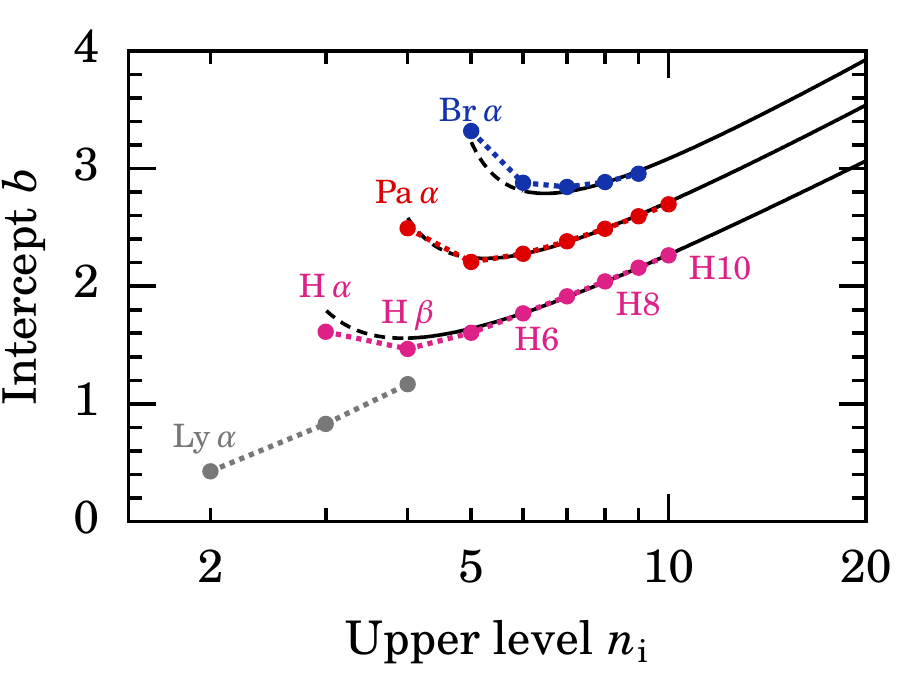}
    \includegraphics[height=\Hoehe]{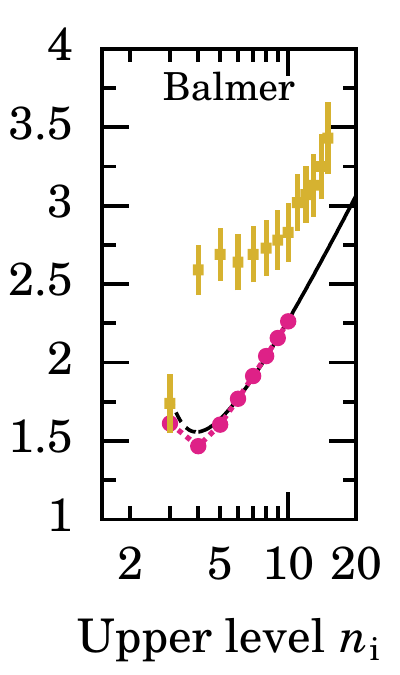}
\end{center}
\caption{%
Fit coefficients of the planetary \Lacc--\Lline relation
(Eq.~(\ref{Gl:LAkkLLinie}))
for  %
Lyman ($\nf=1$) to Brackett ($\nf=4$) lines.
Circles: direct fits of \Lacc--\Lline %
(\citetalias{AMIM21L} and see text).
Black lines: our \textit{fit \vbsrt{of the fit coefficients}} $a$ and $b$
(Eq.~(\ref{Gl:ab})).
Golden squares: \citet{alcal17} for Balmer lines.
}
\label{fig:Koeff}
\end{figure*}

\auskommentiert{
We recall that the hydrogen transitions are determined by the initial and final energy levels \ni and \nf, with \nf defining the series:
Lyman ($\nf=1$),
Balmer ($\nf=2$),
Paschen ($\nf=3$),
Brackett ($\nf=4$),
Pfund ($\nf=5$), etc..
The transitions have energy
$\Delta E(\ni,\nf) = R_\mathrm{y}'\,(1/\nf^2-1/\ni^2)$, where $R_\mathrm{y}'=13.59811$~eV   %
is the modified Rydberg constant taking the electron mass into account \citep{Bearden+Burr1967}.
}

The correlation between \Lacc and each \Lline is
usually written as \citep[e.g.,][]{alcal17}
\begin{equation}
 \label{Gl:LAkkLLinie}
 \log_{10}\left(\Lacc/\Lsun\right) = a \times \log_{10}\left(\Lline/\Lsun\right) + b,
\end{equation}
with $a$ and $b$ the fit coefficients. %
In \citetalias{AMIM21L}, we considered a range of planet masses and accretion rates
and fit the resulting $\Lline(\Lacc)$.
A hydrogen line is defined by the starting and final electron energy levels, \ni and \nf. Since the \citet{aoyama18} models go only up to $n=10$, we did not study lines beyond $\ni=8$ to avoid possible ``edge effects''.
However, higher-level lines are in fact reliable enough (see \S\ref{sec:disc}). Therefore we generate \Lline as in \citetalias{AMIM21L} for an (\Mdot,\,mass) grid for H9, H10, Pa9, Pa10, and Br9, and fit %
Eq.~(\ref{Gl:LAkkLLinie}) to obtain their $a$ and $b$.

Next, we plot the $a$ and $b$ of all lines as a function of \ni and \nf (Fig.~\ref{fig:Koeff}).
We tried different functional forms, fitting the free parameters through \texttt{gnuplot}. %
We used only the $\nf=2$--4 series, excluded the first ($\alpha$) transition of each, and weighted by $1/(\ni-\nf)$. The goal was to have a reliable extrapolation beyond $n\approx10$ (from \citetalias{AMIM21L} or the lines added above). 
An excellent fit is:
\begin{subequations}
\label{Gl:ab}
\begin{align}
 a(\ni,\nf) =& 0.811 - \frac{1}{9.90\nf - 9.5\ni}\label{Gl:a}\\
 b(\ni,\nf) =&  \left[1+1.05\ln(\ni)+ \frac{1}{\ni-\nf}-\frac{1}{\nf}\right] \notag\\
  & \times (1.07+0.0694\nf) -1.41. \label{Gl:b}
\end{align}
\end{subequations}
Even though we excluded %
the $\alpha$ lines, %
their $(a,b)$ predicted by Eq.~(\ref{Gl:ab}) agree closely with the ones from the direct fits. %

Thus, examples of planetary-regime scaling relations extending those of \citetalias{AMIM21L} are:
\begin{align}
 \log_{10}(\Lacc/\LSonne) &= 0.83 \log_{10}(L_{\mathrm{H9}}/\LSonne)  + 2.16  \label{Gl:H9}\\
 \log_{10}(\Lacc/\LSonne) &= 0.82 \log_{10}(L_{\mathrm{H10}}/\LSonne)  + 2.27  \label{Gl:H10}\\
 \log_{10}(\Lacc/\LSonne) &= 0.82 \log_{10}(L_{\mathrm{H11}}/\LSonne)  + 2.37  \label{Gl:H11}\\
 \log_{10}(\Lacc/\LSonne) &= 0.82 \log_{10}(L_{\mathrm{H12}}/\LSonne)  + 2.47  \label{Gl:H12}\\
 \log_{10}(\Lacc/\LSonne) &= 0.82 \log_{10}(L_{\mathrm{H13}}/\LSonne)  + 2.56  \label{Gl:H13}\\
 \log_{10}(\Lacc/\LSonne) &= 0.82 \log_{10}(L_{\mathrm{H14}}/\LSonne)  + 2.64  \label{Gl:H14}\\
 \log_{10}(\Lacc/\LSonne) &= 0.82 \log_{10}(L_{\mathrm{H15}}/\LSonne)  + 2.72   \label{Gl:H15}\\
 \log_{10}(\Lacc/\LSonne) &= 0.82 \log_{10}(L_{\mathrm{H16}}/\LSonne)  + 2.80,  \label{Gl:H16} %
\end{align}
\begin{align}
 \log_{10}(\Lacc/\LSonne) &= 0.83 \log_{10}(L_{\mathrm{Pa9}}/\LSonne)  + 2.60  \label{Gl:Pa9}\\
 \log_{10}(\Lacc/\LSonne) &= 0.83 \log_{10}(L_{\mathrm{Pa10}}/\LSonne)  + 2.72  \label{Gl:Pa10}\\
 \log_{10}(\Lacc/\LSonne) &= 0.82 \log_{10}(L_{\mathrm{Pa11}}/\LSonne)  + 2.82  \label{Gl:Pa11}\\
 \log_{10}(\Lacc/\LSonne) &= 0.82 \log_{10}(L_{\mathrm{Pa12}}/\LSonne)  + 2.92  \label{Gl:Pa12}\\
 \log_{10}(\Lacc/\LSonne) &= 0.82 \log_{10}(L_{\mathrm{Pa13}}/\LSonne)  + 3.01,  \label{Gl:Pa13} %
\end{align}
\begin{align}
 \log_{10}(\Lacc/\LSonne) &= 0.83 \log_{10}(L_{\mathrm{Br9}}/\LSonne)  + 2.98  \label{Gl:Br9}\\
 \log_{10}(\Lacc/\LSonne) &= 0.83 \log_{10}(L_{\mathrm{Br10}}/\LSonne)  + 3.08  \label{Gl:Br10}\\
 \log_{10}(\Lacc/\LSonne) &= 0.83 \log_{10}(L_{\mathrm{Br11}}/\LSonne)  + 3.19  \label{Gl:Br11}\\
 \log_{10}(\Lacc/\LSonne) &= 0.82 \log_{10}(L_{\mathrm{Br12}}/\LSonne)  + 3.29, \label{Gl:Br12}%
\end{align}
simply
evaluating Eq.~(\ref{Gl:ab}) for $\nf=2$--4 and several \ni as examples.
Where $(a,b)$ from direct fits are available, the coefficients match excellently.
Especially since
the $(a,b)$ capture only an average relation,
Eq.~(\ref{Gl:ab}) can also be used for the Pfund or other series ($\nf\geqslant5$).

For error propagation, the errorbar on \Lacc is $\sigma\approx0.3$~dex as for the other lines (Appendix~A of \citetalias{AMIM21L}).
The \texttt{species} toolkit includes
Eqs.~(\ref{Gl:H9})--(\ref{Gl:Br12}) %
(\citealp{stolker20}; ``Emission line'' tutorial at \url{https://species.readthedocs.io}).

\section{Discussion}
 \label{sec:disc}

A few points deserve discussion:
\begin{enumerate}
    \item \textit{Why does it work?}---The lines originate from the shock-heated gas below the shock.
    There, the cooling is slower than or comparable to the electron transitions from low-$n$ levels. %
    This makes non-equilibrium calculations necessary \citep{aoyama18}. However, contrary to the transitions from low-$n$ levels that have a large energy difference, the transitions between high-$n$ levels are faster than the cooling by more than orders of magnitudes.
    Thus, highly-excited-hydrogen abundances follow the thermal (Boltzmann) distribution, which is monotonic, essentially exponential. At velocities and densities relevant here \citep{Aoyama+2020}, electrons with $n \gtrsim 7$ are thermally equilibrated.
    Correspondingly, the line strength of transitions from those $n$ is a simple function. The limiting behaviour seen in Fig.~\ref{fig:Koeff} reflects this.

    \item \textit{Domain of validity}---All slopes
    in all series  %
    converge to $a\approx0.82$, whereas the intercepts are $b\approx2$--3, growing roughly logarithmically
    (Eq.~(\ref{Gl:ab})). This increase of $b$ reflects the intuitive result that a diminishing fraction of \Lacc goes into lines of higher energy within a series ($\Lline/\Lacc\sim10^{-b}$ since $a\sim1$). Formally, integrating \Lline from \nf to infinity diverges, but actually Eq.~(\ref{Gl:ab}) will hold only below a maximum \ni. The highest observable \ni is likely small than this anyway.
    
    \item \textit{Compared to stars?}---In the CTTS relations of \citet{alcal17}, $a_\star\approx1.0$--1.1 while $b_\star\approx2.8$--3.6, also increasing with \ni (Fig.~\ref{fig:Koeff}).
    At \Lline values where the stellar and the planetary regimes somewhat overlap, this leads to a major, 1--4~dex discrepancy in \Lacc between the two approaches \citepalias{AMIM21L}. In the planetary-shock case, a much smaller fraction of the kinetic energy is converted into line luminosity.

   \item \textit{Physical context}---The fits apply to shock emission, whether from magnetospheric or purely hydrodynamic accretion onto the surface of a planet or its CPD.
   However, no emission from accretion columns themselves is considered \citepalias{AMIM21L}. Constraining their %
   temperature structure observationally (\citealp{petrov14}) or theoretically would enable estimating their contribution.
   
   \item \textit{Which way forward?}---The model of \citet{aoyama18} could be extended to calculate directly hydrogen lines involving $n>10$ electrons, but for the purposes of \Lacc--\Lline relations our extrapolations should more than suffice. Adding He and metal lines would  matter more. At preshock velocities $v_0\gtrsim200$~km\,s$^{-1}$, they and the hydrogen continuum carry a sizeable fraction of the energy \citep{Aoyama+2020}, and have been detected 
   (\citealp{eriksson20,zhou21,betti22c,betti22b,ringqvist23}).
   
\end{enumerate}

\section{Summary}

In \citetalias{AMIM21L}, we derived \Lacc--\Lline scalings for accreting planets. These scalings should replace uncalibrated extrapolations of the stellar relations down to low line luminosities, which can yield unphysical results. Here, we fit the \textit{fit coefficients} for the Balmer, Paschen, and Brackett series and extrapolated them to higher-level lines. Thanks to the simplicity of the hydrogen atom, we argued that this estimates well the results of detailed calculations, also for higher series. The underlying scenario is a shock in which hydrogen lines carry most of the energy.
Thus this extension is valid in the same domain of preshock conditions as the main model: for preshock number densities $n_0\sim10^9$--$10^{14}$~cm$^{-3}$ and preshock velocities up to $v_0\approx200$~km\,s$^{-1}$ \citep{Aoyama+2020}.  %
Eq.~(\ref{Gl:ab}) can be used to generate scalings easily, with examples in Eqs.~(\ref{Gl:H9})--(\ref{Gl:Br12}).
These extended relations allow estimates of the accretion rate of planetary-mass objects from even more hydrogen lines than up to now. %

\begin{acknowledgments}
{\small
\vbsrt{We thank Jun Hashimoto, Carlo Manara, and Sarah Betti for useful comments,
Tomas Stolker for adding the fits to \texttt{species},
and Janice Sexton at AAS for her kind help with the publishing process.}
G-DM acknowledges support from the DFG priority program SPP 1992 ``Exploring the Diversity of Extrasolar Planets'' (MA~9185/1),
and from the Swiss National Science Foundation (SNSF), 
grant
200021\_204847 ``PlanetsInTime''.
YA acknowledges support from the National Key R\&D Program of China (No.~2019YFA0405100).
Parts of this work have been carried out within the framework of the NCCR PlanetS supported by the SNSF.
}
\end{acknowledgments}

\bibliography{Gwnffrw_ohni}{}

\begin{thebibliography}{}
\providecommand\natexlab[1]{#1}
\providecommand\JournalTitle[1]{#1}

\bibitem[{{Alcal{\'a}} {et~al.}(2022){Alcal{\'a}}, {Cupani}, {Evans},
  {Franchini}, \& {Nisini}}]{alcal22}
{Alcal{\'a}}, J.~M., {Cupani}, G., {Evans}, C.~J., {Franchini}, M., \&
  {Nisini}, B. 2022,
  \href{http://dx.doi.org/10.1007/s10686-022-09832-1}{\JournalTitle{Experimental
  Astronomy}}, \href{http://arxiv.org/abs/2203.15581}{{\sffamily
  arXiv:2203.15581}}

\bibitem[{{Alcal{\'a}} {et~al.}(2017){Alcal{\'a}}, {Manara}, {Natta}, {Frasca},
  {Testi}, {Nisini}, {Stelzer}, {Williams}, {Antoniucci}, {Biazzo}, {Covino},
  {Esposito}, {Getman}, \& {Rigliaco}}]{alcal17}
{Alcal{\'a}}, J.~M., {Manara}, C.~F., {Natta}, A., {et~al.} 2017,
  \href{http://dx.doi.org/10.1051/0004-6361/201629929}{\JournalTitle{\aap},
  600, A20}

\bibitem[{{Aoyama} \& {Ikoma}(2019)}]{Aoyama+Ikoma2019}
{Aoyama}, Y., \& {Ikoma}, M. 2019,
  \href{http://dx.doi.org/10.3847/2041-8213/ab5062}{\JournalTitle{\apjl}, 885,
  L29}

\bibitem[{{Aoyama} {et~al.}(2018){Aoyama}, {Ikoma}, \& {Tanigawa}}]{aoyama18}
{Aoyama}, Y., {Ikoma}, M., \& {Tanigawa}, T. 2018,
  \href{http://dx.doi.org/10.3847/1538-4357/aadc11}{\JournalTitle{\apj}, 866,
  84}

\bibitem[{{Aoyama} {et~al.}(2021){Aoyama}, {Marleau}, {Ikoma}, \&
  {Mordasini}}]{AMIM21L}
{Aoyama}, Y., {Marleau}, G.-D., {Ikoma}, M., \& {Mordasini}, C. 2021,
  \href{http://dx.doi.org/10.3847/2041-8213/ac19bd}{\JournalTitle{\apjl}, 917,
  L30}

\bibitem[{{Aoyama} {et~al.}(2020){Aoyama}, {Marleau}, {Mordasini}, \&
  {Ikoma}}]{Aoyama+2020}
{Aoyama}, Y., {Marleau}, G.-D., {Mordasini}, C., \& {Ikoma}, M. 2020,
  \JournalTitle{arXiv e-prints}, arXiv:2011.06608

\bibitem[{{Betti} {et~al.}(2022{\natexlab{b}}){Betti}, {Follette},
  {Ward-Duong}, {Aoyama}, {Marleau}, {Bary}, {Robinson}, {Janson}, {Balmer},
  {Chauvin}, \& {Palma-Bifani}}]{betti22b}
{Betti}, S.~K., {Follette}, K.~B., {Ward-Duong}, K., {et~al.}
  2022{\natexlab{a}},
  \href{http://dx.doi.org/10.3847/2041-8213/ac85ef}{\JournalTitle{\apjl}, 935,
  L18}

\bibitem[{{Betti} {et~al.}(2022{\natexlab{a}}){Betti}, {Follette},
  {Ward-Duong}, {Aoyama}, {Marleau}, {Bary}, {Robinson}, {Janson}, {Balmer},
  {Chauvin}, \& {Palma-Bifani}}]{betti22c}
{Betti}, S.~K., {Follette}, K.~B., {Ward-Duong}, K., {et~al.}
  2022{\natexlab{b}},
  \href{http://dx.doi.org/10.3847/2041-8213/aca331}{\JournalTitle{\apjl}, 941,
  L20}

\bibitem[{{Dong} {et~al.}(2021){Dong}, {Jiang}, \& {Armitage}}]{dong21}
{Dong}, J., {Jiang}, Y.-F., \& {Armitage}, P.~J. 2021,
  \href{http://dx.doi.org/10.3847/1538-4357/ac1941}{\JournalTitle{\apj}, 921,
  54}

\bibitem[{{Eriksson} {et~al.}(2020){Eriksson}, {Asensio Torres}, {Janson},
  {Aoyama}, {Marleau}, {Bonnefoy}, \& {Petrus}}]{eriksson20}
{Eriksson}, S.~C., {Asensio Torres}, R., {Janson}, M., {et~al.} 2020,
  \href{http://dx.doi.org/10.1051/0004-6361/202038131}{\JournalTitle{\aap},
  638, L6}

\bibitem[{{Haffert} {et~al.}(2019){Haffert}, {Bohn}, {de Boer}, {Snellen},
  {Brinchmann}, {Girard}, {Keller}, \& {Bacon}}]{Haffert+2019}
{Haffert}, S., {Bohn}, A., {de Boer}, J., {et~al.} 2019,
  \href{http://dx.doi.org/10.1038/s41550-019-0780-5}{\JournalTitle{Nat.\
  Astron.}, 3, 749}

\bibitem[{{Komarova} \& {Fischer}(2020)}]{Komarova+Fischer2020}
{Komarova}, O., \& {Fischer}, W.~J. 2020,
  \href{http://dx.doi.org/10.3847/2515-5172/ab67bb}{\JournalTitle{RNAAS}, 4, 6}

\bibitem[{{Petrov} {et~al.}(2014){Petrov}, {Gahm}, {Herczeg}, {Stempels}, \&
  {Walter}}]{petrov14}
{Petrov}, P.~P., {Gahm}, G.~F., {Herczeg}, G.~J., {Stempels}, H.~C., \&
  {Walter}, F.~M. 2014,
  \href{http://dx.doi.org/10.1051/0004-6361/201424374}{\JournalTitle{\aap},
  568, L10}

\bibitem[{{Rigliaco} {et~al.}(2012){Rigliaco}, {Natta}, {Testi}, {Randich},
  {Alcal{\`a}}, {Covino}, \& {Stelzer}}]{rigliaco12}
{Rigliaco}, E., {Natta}, A., {Testi}, L., {et~al.} 2012,
  \href{http://dx.doi.org/10.1051/0004-6361/201219832}{\JournalTitle{\aap},
  548, A56}

\bibitem[{{Ringqvist} {et~al.}(2023){Ringqvist}, {Viswanath}, {Aoyama},
  {Janson}, {Marleau}, \& {Brandeker}}]{ringqvist23}
{Ringqvist}, S.~C., {Viswanath}, G., {Aoyama}, Y., {et~al.} 2023,
  \href{http://dx.doi.org/10.1051/0004-6361/202245424}{\JournalTitle{\aap},
  669, L12}

\bibitem[{{Stolker} {et~al.}(2020){Stolker}, {Quanz}, {Todorov}, {K{\"u}hn},
  {Molli{\`e}re}, {Meyer}, {Currie}, {Daemgen}, \& {Lavie}}]{stolker20}
{Stolker}, T., {Quanz}, S., {Todorov}, K., {et~al.} 2020,
  \href{http://dx.doi.org/10.1051/0004-6361/201937159}{\JournalTitle{\aap},
  635, A182}

\bibitem[{{Szul{\'a}gyi} \& {Ercolano}(2020)}]{szul20}
{Szul{\'a}gyi}, J., \& {Ercolano}, B. 2020,
  \href{http://dx.doi.org/10.3847/1538-4357/abb5a2}{\JournalTitle{\apj}, 902,
  126}

\bibitem[{{Thanathibodee} {et~al.}(2019){Thanathibodee}, {Calvet}, {Bae},
  {Muzerolle}, \& {Hern{\'a}ndez}}]{thanathibodee19}
{Thanathibodee}, T., {Calvet}, N., {Bae}, J., {Muzerolle}, J., \&
  {Hern{\'a}ndez}, R.~F. 2019,
  \href{http://dx.doi.org/10.3847/1538-4357/ab44c1}{\JournalTitle{\apj}, 885,
  94}

\bibitem[{{Wu} {et~al.}(2022){Wu}, {Bowler}, {Sheehan}, {Close}, {Eisner},
  {Best}, {Ward-Duong}, {Zhu}, \& {Kraus}}]{wu22}
{Wu}, Y.-L., {Bowler}, B.~P., {Sheehan}, P.~D., {et~al.} 2022,
  \href{http://dx.doi.org/10.3847/2041-8213/ac6420}{\JournalTitle{\apjl}, 930,
  L3}

\bibitem[{{Zhou} {et~al.}(2021){Zhou}, {Bowler}, {Wagner}, {Schneider}, {Apai},
  {Kraus}, {Close}, {Herczeg}, \& {Fang}}]{zhou21}
{Zhou}, Y., {Bowler}, B.~P., {Wagner}, K.~R., {et~al.} 2021,
  \href{http://dx.doi.org/10.3847/1538-3881/abeb7a}{\JournalTitle{\aj}, 161,
  244}

\end{thebibliography}
\bibliographystyle{yahapj.bst}

\end{document}